# The effect of heavy reflector on neutronic parameters of core

Michal Košťál*[1]; Evžen Losa[1]; Tomáš Czakoj[1]; Martin Schulc[1]; Jan Šimon[1], Vlastimil Juříček[1], Vojtěch Rypar[1], Jana Ulmanová[1], Andrej Trkov[2], Roberto Capote[3]

[1] Research Center Rez Ltd, 250 68 Husinec-Rez 130, Czech Republic
[2] Jožef Stefan Institute, Jamova cesta 39, SI-1000 Ljubljana, Slovenia
[3] Nuclear Data Section, International Atomic Energy Agency, A-1400 Wien, Austria

Email: Michal.Kostal@cvrez.cz
Telephone: +420266172655

Abstract
The reactor baffle is an important component of a nuclear reactor that fixes the location of the fuel assembly in the reactor core. The main types of baffles are called light or heavy. The light baffle has mostly the form of steel plates with outer space filled by water, and the heavy baffle is mostly a forged steel element. Both concepts have advantages as well as disadvantages. In the case of the light baffle, one does not need to solve the issue of void swelling, but the neutron economy is not ideal, while the heavy baffle has a good neutron economy, but void swelling is an issue. This paper deals with the effect of the heavy VVER-1000 baffle on criticality. Criticality was measured using a well-defined core composed of 6 fuel assemblies moved to a simulator of the VVER-1000 internals, which is located at the LR-0 reactor. The experiments confirm the fact that the water filling the cooling channels in the baffle has a strong neutron absorbing effect. The $k_{eff}$ calculated using the ENDF/B-VIII.0 library significantly underpredicts the experiment, whereas calculations using a new evaluation of $^{56}$Fe by the IAEA (INDEN collaboration) give a better agreement. Generally, the presented results are suitable for validation of iron cross sections.

## 1   Introduction

The most common classification of baffles divides them as light or heavy baffles. The light baffle has the form of thin steel plates fixing the core elements and water filling space between such plates and barrel. For example, in the VVER-440 reactor, such a plate has a thickness of 8 mm.

This approach has the advantage of eliminating the issue of void swelling in the baffle Yang et al 2020 caused by the temperature field in baffle Harutyunyan et al 2018. On the other hand, the neutron economy and flux in the reactor pressure vessel deteriorate compared to a heavy reflector Taforeau et al 2019 . In addition to the VVER-440 reactor housing, the light reflector is used also in the current French PWR fleet (Generation II), while the VVER-1000 or EPR uses a heavy reflector Santamarina et al 2008.

The better neutron economy of the heavy reflector comes with the higher back-scattering effect for fast neutrons in the case of iron compared to water. Due to the relatively low cross sections in the higher energy region, the positive effect becomes evident for the higher thickness of steel, while for the thin steel plate followed by water, the absorbing properties dominate and the light reflector acts as a strong absorber Tahara et al 2001 for thermalized neutrons behind the steel plate.

Discrepancies were reported in the $k_{eff}$ calculations of cores with heavy reflectors, which seem to be an effect of the discrepant description of iron in current nuclear data libraries. The aim of the present work was to measure the criticality in benchmark configurations that use light or heavy reflectors using a very well characterized VVER-1000 internal core simulator. As $k_{eff}$ is a basic parameter of criticality safety, it is necessary to validate critical core models with a steel reflector. The experiments are also usable for the validation of the iron cross section because the employed reactor core configuration is very well-defined, and led to the establishment of the reference neutron field in the center of the default geometry with full water reflector (Kostal et al 2020, Trkov 2020).

## 2 Experimental and calculation methods

### 2.1 LR-0 reactor arrangement

The experimental work was carried out at the LR-0 reactor in Rez. Reactor LR-0 is an experimental light-water-moderated zero-power reactor Kostal et al. 2016a. It uses the hexagonal fuel, radially identical to the VVER-1000 fuel (with lattice pitch 12.75 mm). Comparing to the VVER-1000, it is axially shortened to 125 cm. The fuel is dismountable, and thus the experiments focused on fission density determination in each pin can be performed there. Due to the used fuel, the LR-0 was originally designed for experimental research of the VVER-1000 and VVER-440 type reactor cores, spent-fuel storage lattices, and similar benchmark experiments. Criticality control can be done by moving the control rods or by changing the moderator level. In this experimental set, criticality is reached by changing the water level exclusively.

The continuous maximal operating power is 1 kW, in the small, partially flooded cores, the common maximal power used in activation cross section measurement is ~7 W with thermal neutron flux in the core center ~3E7 $n \times cm^{-2} \times s^{-1}$ and fast neutron flux (> 0.1 MeV) ~5E7 $n \times cm^{-2} \times s^{-1}$.

Thanks to the versatile design of the LR-0 fuel supporting structures, several experiments using the same core and different reflectors were carried out, which allowed not only validation of the mathematical description, but also a direct comparison between different reflectors. In total, 5 cases were studied in detail from the point of view of reactivity and power distribution. The first case (Case 1) is a reference, where the core has been characterized by many experimental works in the past (Kostal et al. 2016b, Kostal et al. 2017c, Kostal et al. 2017b, Kostal et al. 2016a). The core in Case 1 consists of only 6 fuel assemblies surrounding a large empty (void) channel, and only a water reflector is placed around this core. Additionally, 6 tubes for reactor instrumentations are positioned around the core (see Figure 1). Case 1A has, similarly to Case 1, 6 fuel assemblies surrounding the large empty channel, but the instrumentation tubes are further from the core.

Case 2 is based on Case 1, but empty channels were placed in 4 assembly positions around the core, increasing neutron leakage from the core in these directions. In Case 3, the core is moved closer to the VVER-1000 heavy steel reflector, near the cooling channels. Case 4 introduces iron fillers into a large and one small cooling channel in the baffle, thus replacing water with low-alloy steel. The core schematics for all cases are depicted in Figure 1, the Case 1A (with channels ~4cm from fuel assemblies) is plotted in Figure 2.

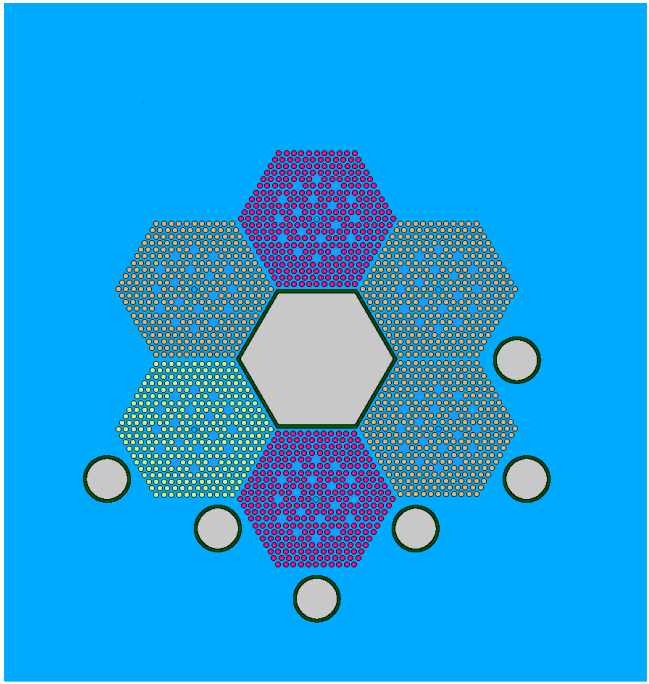
Case 1

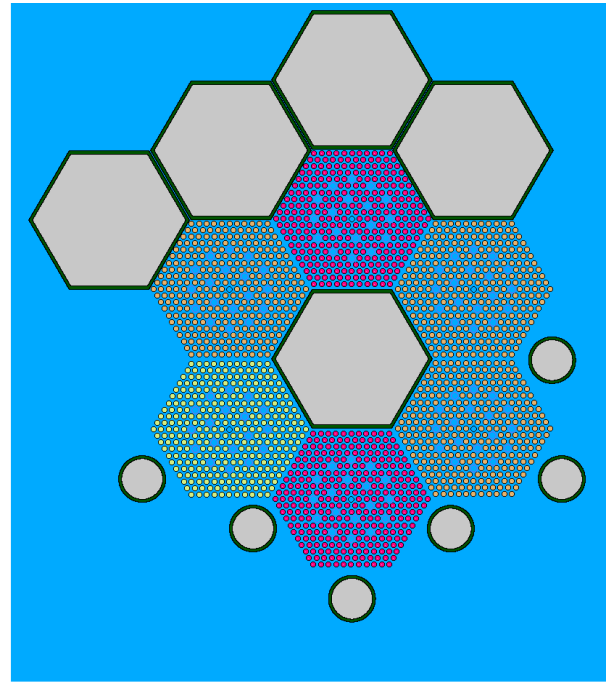
Case 2

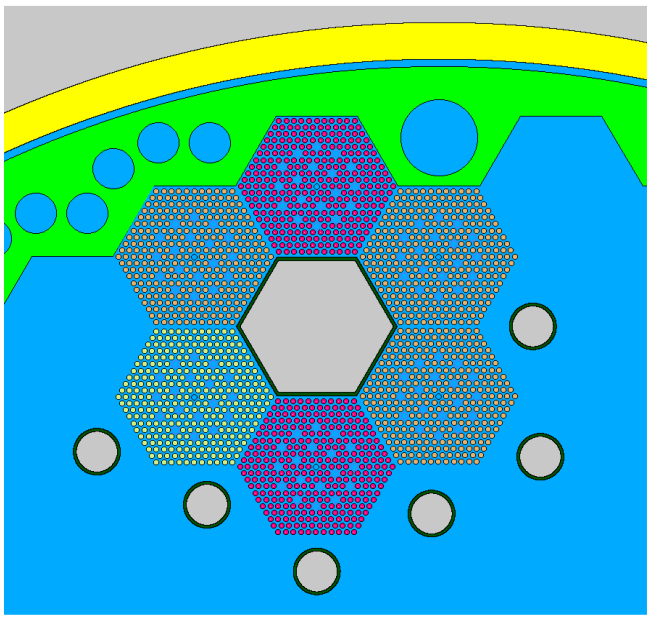
Case 3

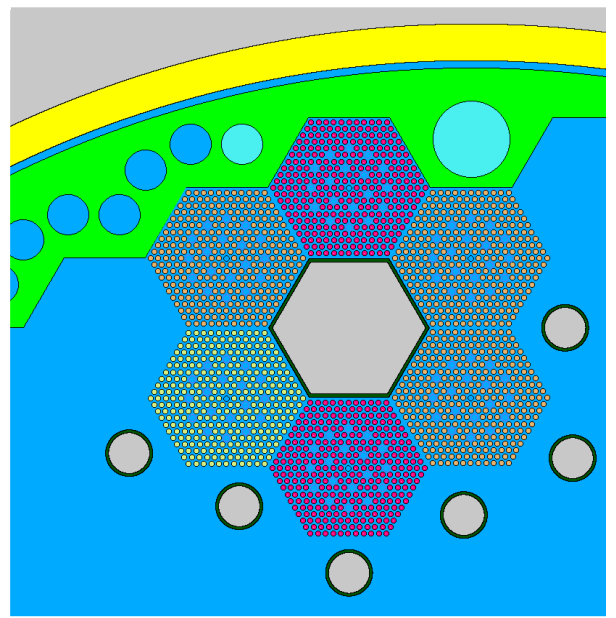
Case 4

Figure 1: The experimental configurations with various reflector designs

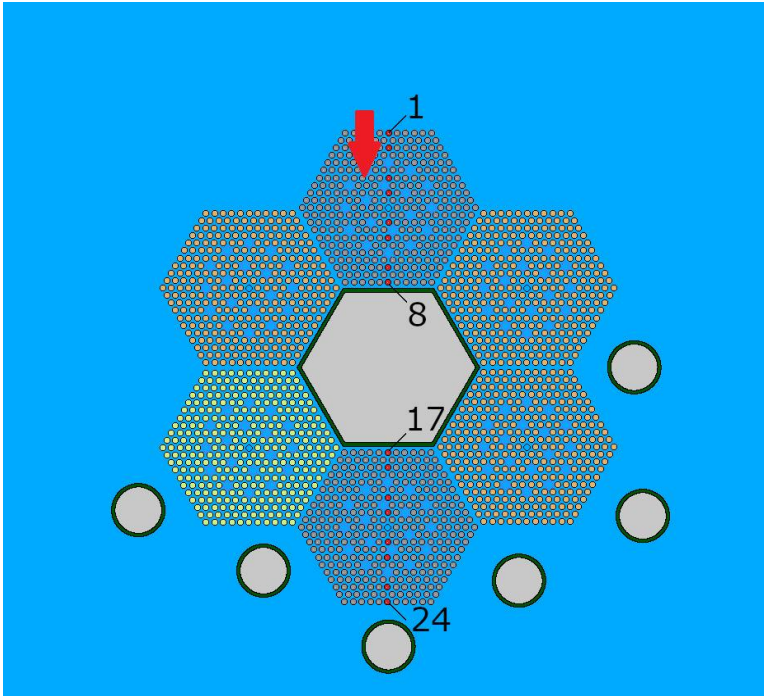

Figure 2: Case 1A with highlighted measured pins and instrumentation channels 4 cm from fuel

## 2.2 Reactivity measurement

Reactivity of the cores for the described experiments is driven exclusively by the water moderator level. In that case moderator level ($H_{cr}$) acts as a critical parameter. If the moderator level slightly exceeds the $H_{cr}$, the one-group asymptotic approximation may be used to express the reactivity (1)

$$\rho(H) = \frac{C}{(H_{cr}+\lambda_z)^2}\left[\frac{1}{(1+\frac{H-H_{cr}}{H_{cr}+\lambda_z})^2}-1\right]; \rho = \frac{k_{eff}-1}{k_{eff}}; k_{eff} = \frac{k_\infty}{1+M^2 B^2} \quad (1)$$

where $C$ is constant; $C = \frac{M^2\pi^2}{k\infty}$; $B^2$ geometry buckling, $M^2$ migration area and $\lambda_z$ is the axial extrapolation length.

For low reactivities, the formula (1) may be expanded into the Taylor series around the $H_{cr}$ as described by expression (2):

$$\rho(H) = f(H, a_1, a_2) = a_1 \cdot (H-a_2) \cdot \left(1 - \frac{3}{2}\frac{H-a_2}{a_2+\lambda_z}\right) \quad (2)$$

where $a_1 = \dfrac{\delta\rho}{\delta H}$ ; $a_2 = H_{Cr}$

Using this formula, critical levels and reactivity coefficients were evaluated from different supercritical states measured using the inverse kinetics method based on the time-dependent response of counters placed in the instrumentation tubes around the reactor core. The total uncertainty of the $H_{cr}$ evaluation is 0.058 cm and includes the technical tolerance and the uncertainty of the level meter calibration.

### 2.3 Fission density measurement

For the complex characterization of the core, not only the reactivity but also the distribution of the fission density distribution in the core (see Figure 2) was studied. The fission density was determined from the gamma activity of the $^{92}$Sr fission product induced during the irradiation experiment (Eq. 3, 4). The gamma activity was determined using the well-characterized HPGe detector Kostal et al 2018. The existence of the validated HPGe calculation model is necessary because in this studied case there is a different efficiency for each pin (see Table 1). It is a consequence of the fact that in a partially flooded core with central void assembly, the axial power profile differs for each pin (see Figure 3). The variations in the detection efficiencies for the pins across each core mainly reflect especially variations in the relative share of the collimated part on the total pin fission density, as well as the contribution of the relative photon transmission through the collimator Kostal et al. 2016b.

Due to the symmetry in the core, the efficiencies of symmetrically located pins are comparable (see Table 2). Similarly, in pins farther from the boundary, the fission profile and following efficiency is comparable. The observed variations from average efficiency reflect the variations in power density profile Kostal et al. 2016b.

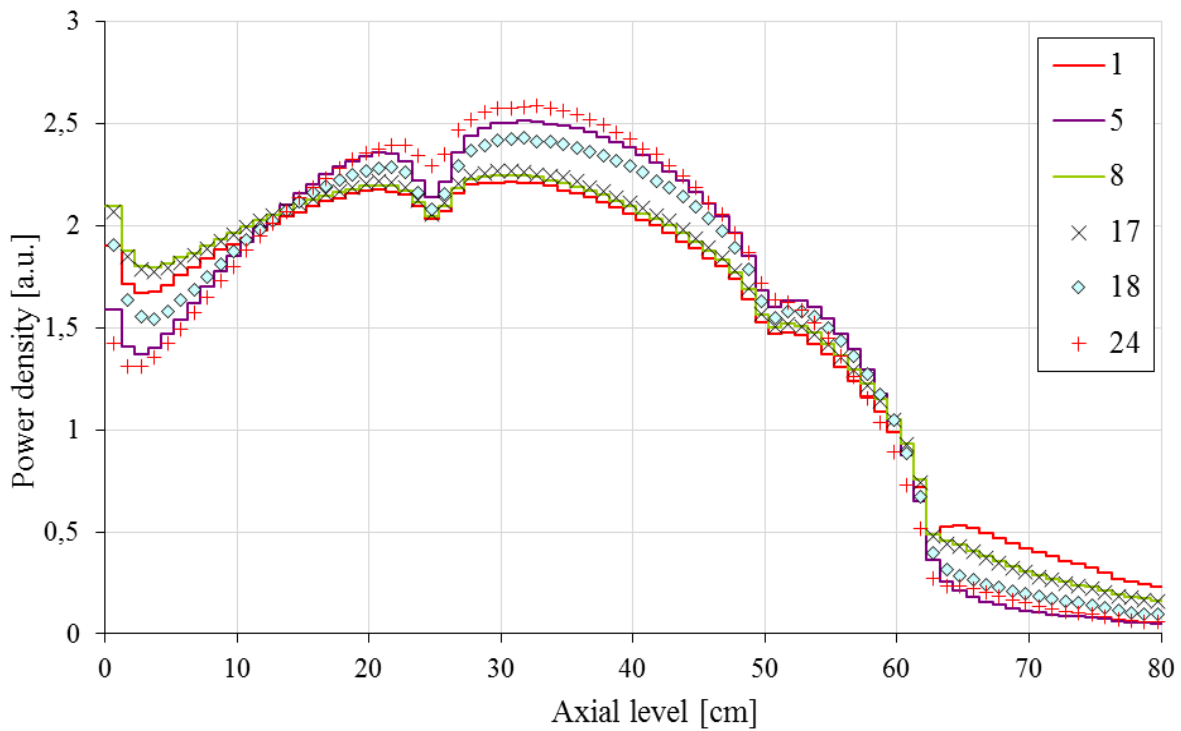

Figure 3: Axial profile in various pins across the cut in the C2 core with 4 void assemblies. Pin numbering is visualized in Figure 2

$$F_j^i = \frac{NPA_j^i(t)}{\eta^i \cdot \varepsilon^i \cdot \lambda^i} \cdot \frac{1}{N^i(t)} = NPA_j^i(0) \cdot \frac{1}{\eta^i \cdot \varepsilon^i \cdot \lambda^i \cdot N^i(0)} \quad (3)$$

$$NPA_j^i(0) = NPA_j^i(t) \cdot \frac{1}{e^{-\lambda^i \cdot t}} \cdot \frac{\lambda^i \cdot \Delta T}{1 - e^{-\lambda^i \cdot \Delta T}} \quad (4)$$

Where:

$F_j^i$ fission rate determined via the i-th nuclei and j-th pin;

$N^i(t)$ calculated number of observed nuclei in fuel pin when 1 fission/s occurs, in time t after irradiation end;

$NPA_j^i(t)$ measured Net Peak Area j-th pin of the observed nuclei i and selected peak,

$\lambda^i$ decay constant of selected nuclide;

$\eta^i$ efficiency of HPGe for the selected gamma line of the i-th nuclide;

$\varepsilon^i$ gamma branching ratio of the selected peak from observed nuclei i;

$t$ start j-th pin measurement;

$\Delta T$ length of j-th pin HPGe measurement

Table 1.: Efficiencies of measuring geometries for pins across the core in various core arrangements

| Pin | C4 | C3 | C2 | C1 |
|---|---|---|---|---|
| 1 | 2.95E-04 | 2.99E-04 | 2.44E-04 | 3.16E-04 |
| 2 | 2.97E-04 | 3.01E-04 | 2.64E-04 | 3.11E-04 |
| 3 | 3.00E-04 | 3.03E-04 | 2.76E-04 | 3.09E-04 |
| 4 | 2.99E-04 | 3.03E-04 | 2.77E-04 | 3.08E-04 |
| 5 | 2.97E-04 | 3.00E-04 | 2.76E-04 | 3.04E-04 |
| 6 | 2.95E-04 | 2.98E-04 | 2.74E-04 | 3.00E-04 |
| 7 | 2.82E-04 | 2.84E-04 | 2.61E-04 | 2.85E-04 |
| 8 | 2.72E-04 | 2.69E-04 | 2.44E-04 | 2.67E-04 |
| 17 | 2.72E-04 | 2.70E-04 | 2.47E-04 | 2.68E-04 |
| 18 | 2.83E-04 | 2.86E-04 | 2.61E-04 | 2.82E-04 |
| 19 | 2.98E-04 | 3.01E-04 | 2.74E-04 | 2.97E-04 |
| 20 | 3.01E-04 | 3.05E-04 | 2.76E-04 | 3.00E-04 |
| 21 | 3.04E-04 | 3.08E-04 | 2.77E-04 | 3.05E-04 |
| 22 | 3.05E-04 | 3.10E-04 | 2.76E-04 | 3.07E-04 |
| 23 | 3.05E-04 | 3.09E-04 | 2.76E-04 | 3.05E-04 |
| 24 | 3.12E-04 | 3.13E-04 | 2.76E-04 | 3.06E-04 |

Table 2.: Average standard pin efficiency and difference from used pins

|    | C4    | C3     | C2     | C1     |
|----|-------|--------|--------|--------|
| 1  | -1.5% | -1.3%  | -11.6% | 4.0%   |
| 2  | -1.1% | -0.9%  | -4.3%  | 2.3%   |
| 3  | 0.0%  | -0.1%  | 0.1%   | 1.7%   |
| 4  | -0.2% | -0.3%  | 0.5%   | 1.3%   |
| 5  | -0.9% | -1.0%  | 0.1%   | 0.1%   |
| 6  | -1.7% | -1.9%  | -0.7%  | -1.1%  |
| 7  | -6.0% | -6.5%  | -5.3%  | -6.2%  |
| 8  | -9.4% | -11.4% | -11.3% | -12.0% |
| 17 | -9.3% | -11.1% | -10.5% | -11.9% |
| 18 | -5.5% | -5.8%  | -5.3%  | -7.1%  |
| 19 | -0.7% | -0.7%  | -0.7%  | -2.3%  |
| 20 | 0.3%  | 0.4%   | 0.1%   | -1.2%  |
| 21 | 1.5%  | 1.6%   | 0.5%   | 0.5%   |
| 22 | 1.8%  | 2.0%   | 0.1%   | 1.0%   |
| 23 | 1.9%  | 1.9%   | 0.1%   | 0.6%   |
| 24 | 4.1%  | 3.3%   | 0.1%   | 0.7%   |

### 2.4 Gamma spectrometry of activation detectors

In the experiment, where the core is surrounded by water (Case 1A), the power profile was determined in absolute numbers. The absolute flux was determined by means of the flux method, where the flux was determined from reaction rates of pure Mn monitor Trkov 2020. The Mn monitors were placed in the reference position in the central hexagonal channel (see Kostal et al 2020).

The efficiency of the used measuring geometries was determined by calculation because there is a validated model of HPGe for activated foils measurements Kostal et al. 2017b. The details dealing with the used methodology can be found in Boson et al. 2008 and Dryak and Kovar 2006. The calculation model allows the determination of the efficiency as well as the coincidence summing factors at the same time.

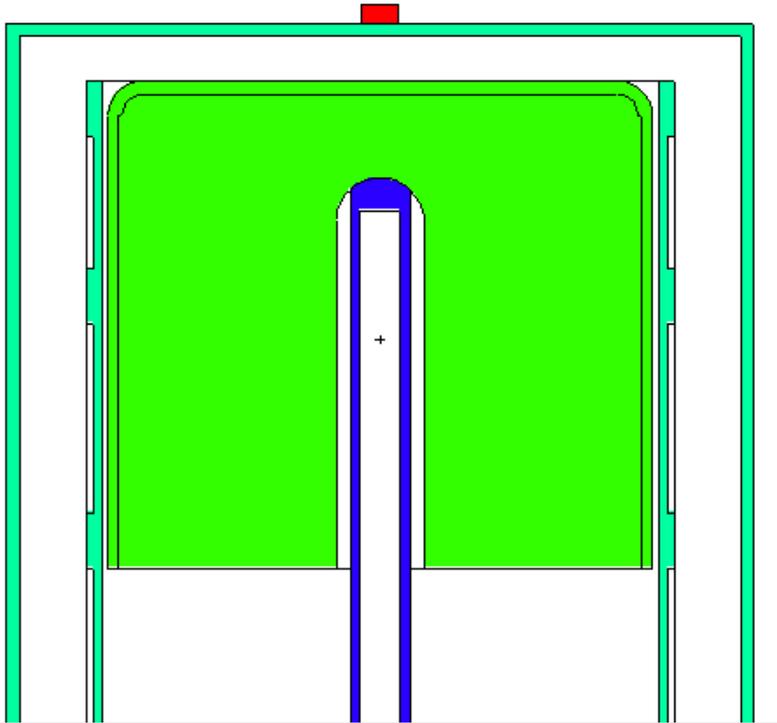

Figure 4: Measured geometry of natural Mn foils

### 2.5 Calculational methods

The calculations were performed with the MCNP6 code [Goorley et al. 2012](#) and various nuclear data libraries: ENDF/B-VIII.0 [Brown et al 2018](#), ENDF/B-VII.1 [Chadwick et al. 2011](#), ENDF/B-VI.2, JEFF-3.3 [Koning et al. 2010](#), JENDL-4 [Shibata et al. 2011](#), ROSFOND-2010 [Zabrodskaya et al. 2007](#) and CENDL-3.1 [Ge et al 2010](#). Testing evaluation data of iron was obtained from IAEA (INDEN collaboration downloaded from [https://www-nds.iaea.org/INDEN/data/fe56e80X29r41_ENDF.zip](https://www-nds.iaea.org/INDEN/data/fe56e80X29r41_ENDF.zip)), more details about these data are given below in chapter 2.6.

The reactor model was compiled using the experimentally measured critical water level. Thus the rate of disagreement is simply the difference from unity because the critical water level is determined for the critical configuration, therefore the experimental $k_{eff}$ is 1. Uncertainties in the modeled system are based on manufacturing tolerances for the fuel and structural parts of the core, uncertainty in critical parameter measurement, and uncertainties in the composition of materials in the reactor. Uncertainties are estimated by calculation and are about 140 pcm. In relative comparison between the two cases without manipulation with fuel is only about 20 pcm (between cases 3 and 4 and Case 1 and Case 1A). More details concerning uncertainty analysis can be found in [Kostal et al 2018b](#).

Calculations have been performed in the critical mode of the MCNP6.2 code using 10000 generations and 40000 neutrons in one generation.

Sensitivity was determined using the SCALE6.2/TSUNAMI-3D code [Rearden and Jessee 2017](#). Calculations were performed using the Iterated Fission Probability method [Perfetti et al. 2016](#) and the ENDF/B-VII.1 continuous energy library. Each case was calculated with a different number of neutrons per generation and a different number of active generations, which were necessary for sufficient result convergence.

## 2.6 Nuclear data

The INDEN (IAEA) Fe-56,57 evaluations were developed to correct problems identified in the fast region of the ENDF/B-VIII.0 iron evaluation (see Refs. [Herman et al 2018, Brown et al 2018, Chadwick et al 2018]. In particular, a 30% underestimation of the fast neutron leakage was identified and mostly associated with an overestimation of the inelastic cross section above 2 MeV. New evaluation is available from the INDEN project webpage https://www-nds.iaea.org/INDEN/. A new evaluation of chromium isotopes was also recently published by INDEN collaboration ( Nobre et al 2021).

## 3 Results

### 3.1 Reactivity

The critical parameter, namely the moderator critical level, was determined for each studied case (see Table 3). When the water reflector is replaced by the VVER-1000 internals simulator, the critical level is slightly increasing. The introduced reactivity is about 60 pcm, which means that the water reflector has a better neutron economy than the VVER-1000.

When two most effective cooling channels were filled by steel insertions (Case 4), the critical level significantly decreases. The reactivity obtained comparing with water reflector case (Case 1) is about 290 pcm. A similar manner where a steel reflector has a better neutron economy than a water reflector was reported by Tahara et al 2001.

Using the experimental critical moderator level, the calculation model of the reactor was assembled in the MCNP6 code. A comparison of the cases with the water reflector or partially non-reflecting shows a good agreement of the calculated value with the experimental value of 1, because the core is critical.

When the core is next to the VVER-1000 internals simulator, the disagreement of about 230 pcm can be observed. It should be noted that when the two most influencing cooling channels are filled by the low-alloy steel insertions, the rate of disagreement is slightly lower.

The effect of instrumentation tube distances from the core is also visible (Case 1 and Case 1A). According to the expectation, when channels are further from the core, the reflection properties get better, and the critical level is lower compared to the case with closer channels.

Table 3.: Critical levels and following keff calculated with MCNP6 and ENDF-B/VIII.0

|   | | Hcr | unc. | ENDF/B-VIII.0 |
| --- | --- | --- | --- | --- |
| Water reflected core and channels displaced 4cm from fuel | Case 1A | 538.48 | 0.17 | 1.00002 |
| Water reflected core | Case 1 | 549.39 | 0.41 | 0.99987 |
| 4 void assemblies | Case 2 | 621.78 | 0.17 | 0.99969 |
| Near baffle | Case 3 | 540.70 | 0.72 | 0.99774 |
| Near baffle + steel in channels | Case 4 | 527.47 | 0.04 | 0.99797 |

The calculations performed using various nuclear data libraries are listed in Table 4. For general-purpose evaluated libraries, the best agreement was reached for the case with water reflector and case with the partially non-reflected core (Cases 1A,1,2). ENDF/B-VIII.0 shows the best agreement ($\Delta k_{eff}$ = -15pcm average), but the performance of JEFF-3.3, JENDL-4, and ROSFOND-2010 is similar within 100 pcm. ENDF/B-VI.2 and CENDL-3.1 underestimate criticality for those cases.

When the VVER-1000 internals simulator is adjoining the core, the agreement worsens by about 200 pcm, especially for the Case 3, with JEFF-3.3 showing the best agreement. The highest decrease is in CENDL-3.1 and the lowest one in ENDF/B-VI.2. The rate of disagreement in the case of ENDF/B-VI.2 is comparable with Kostal et al 2013. Tahara et al 2001 with MVP code and ENDF/B-VI.2 reported $k_{eff}$ 0.99487 while in geometry with Fe reflector $k_{eff}$ was 0.99569. The presented results determined with MCNP6 and ENDF/B-VI.2 demonstrate an opposite trend, being 0.99281 for water reflector (Case 1) while for VVER-1000 heavy baffle reflector calculated $k_{eff}$ worsens to 0.99085 in the Case 3. Adding iron to the cooling channels generally leads to the improvement of the calculation vs. experiment agreement.

New INDEN evaluations for Fe and Cr isotopes were tested only in combination with ENDF/B-VIII.0 data and do not improve the results. It should be noted that oxygen absorption has a very large impact on all presented results. Criticality spread is reduced significantly when we switch from O-16 ENDF/B-VIII.0 evaluation (which features high absorption in oxygen for fast neutrons) to ENDF/B-VII.1 evaluation for O-16.

Table 4.: Keff calculated with MCNP6 and various data libraries

|  | ENDF/B-VIII.0 | ENDF/B-VII.1 | ENDF/B-VI.2 | JEFF-3.3 | JENDL-4 | ROSFON D-2010 | CENDL-3.1 |
|---|---|---|---|---|---|---|---|
| Case 1A | 1.00002 | 1.00075 | 0.99281 | 1.00107 | 1.00035 | 1.00015 | 0.99946 |
| Case 1 | 0.99987 | 1.00072 | 0.99266 | 1.00092 | 1.00032 | 1.00007 | 0.99916 |
| Case 2 | 0.99969 | 1.00032 | 0.99253 | 1.00083 | 1.00001 | 0.99968 | 0.99894 |
| Case 3 | 0.99774 | 0.99838 | 0.99085 | 0.99895 | 0.99818 | 0.99769 | 0.99657 |
| Case 4 | 0.99797 | 0.99873 | 0.99126 | 0.99931 | 0.99843 | 0.99807 | 0.99678 |
| $<\Delta k_{eff}>$ | 0.00095 | 0.00094 | 0.00798 | 0.00091 | 0.00081 | 0.00096 | 0.00182 |

Table 5.: Keff calculated with MCNP6 and ENDF/B-VIII with changed evaluations for iron, chromium and oxygen

|  | Fe INDEN | Fe, Cr INDEN | Fe INDEN; O ENDF/B-VII.1 | Fe, Cr INDEN; O ENDF/B-VII.1 |
|---|---|---|---|---|
| Case 1A | 1.00004 | 0.99981 | 1.00134 | 1.00102 |
| Case 1 | 0.99985 | 0.99973 | 1.00122 | 1.0011 |
| Case 2 | 0.99965 | 0.99945 | 1.00099 | 1.00073 |
| Case 3 | 0.99772 | 0.99734 | 0.99897 | 0.99857 |
| Case 4 | 0.99803 | 0.99758 | 0.99936 | 0.99897 |
| $<\Delta k_{eff}>$ | 0.00096 | 0.00122 | 0.00104 | 0.00106 |

### 3.2 Sensitivities to selected reactions

The addition of steel reflector in Cases 3 and 4 significantly changes the good agreement in calculated $k_{eff}$ comparing to the case when the core is reflected by water or even partly non-reflected. This raises a question on which materials and reactions might introduce such discrepancy into the calculational model. Due to this fact, the set of sensitivity calculations was performed. It results are presented in Table 6 and Table 7.

The steel components play a significant role in all studied cores because a large amount of steel forms fuel structural components. But when it acts as a fuel component (like a spacing grid) or fuel support structures, it has a negative effect on criticality because capture significantly prevails over the scattering. In the case of the baffle, where it directly adjoins to fuel, the situation is opposite, and the effect of neutron scatter (especially elastic) prevails over the capture.

Comparison of Table 6 and Table 7 with criticality results show that better agreement of calculation with experiment can be found in Case 4, where the iron isotopes in baffle effect play more important role that in Case 3.

The energy-dependent sensitivity is plotted for the case of baffle and most dominant isotopes, namely $^{56}$Fe isotope in Figure 5 and $^{52}$Cr in Figure 6.

Table 6.: Sensitivity of steel components in reference core and in core near baffle geometry

|      | Case 1 (in whole core) | | | Case 3 (in whole core) | | | Case 3 (effect of baffle) | | |
|------|-------|--------|--------|-------|--------|--------|-------|--------|--------|
|      | (n,n) | (n,n') | (n,γ)  | (n,n) | (n,n') | (n,γ)  | (n,n) | (n,n') | (n,γ)  |
| $^{56}$Fe | 302 | 99 | -1560 | 942 | 204 | -1793 | 613 | 106 | -216 |
| $^{52}$Cr | 64  | 22 | -137  | 211 | 42  | -159  | 143 | 20  | -20  |
| $^{58}$Ni | 65  | 7  | -266  | 199 | 14  | -307  | 132 | 7   | -38  |
| $^{54}$Fe | 22  | 4  | -88   | 74  | 8   | -102  | 49  | 4   | -13  |
| $^{60}$Ni | 16  | 4  | -59   | 49  | 7   | -68   | 33  | 3   | -9   |
| $^{57}$Fe | 6   | 4  | -35   | 20  | 11  | -40   | 13  | 7   | -6   |
| $^{53}$Cr | 13  | 4  | -321  | 51  | 7   | -364  | 32  | 4   | -43  |

Table 7.: Sensitivity of steel components in Case 4

|      | Case 4 (in whole core) | | | Case 4 (effect of baffle) | | |
|------|-------|--------|--------|-------|--------|--------|
|      | (n,n) | (n,n') | (n,γ)  | (n,n) | (n,n') | (n,γ)  |
| $^{56}$Fe | 1160 | 235 | -1820 | 812 | 138 | -240 |
| $^{52}$Cr | 225  | 42  | -159  | 153 | 20  | -21  |
| $^{58}$Ni | 225  | 14  | -307  | 153 | 7   | -39  |
| $^{54}$Fe | 99   | 10  | -105  | 74  | 6   | -15  |
| $^{60}$Ni | 52   | 7   | -68   | 36  | 4   | -9   |
| $^{57}$Fe | 29   | 14  | -42   | 21  | 9   | -7   |
| $^{53}$Cr | 56   | 8   | -364  | 37  | 4   | -43  |

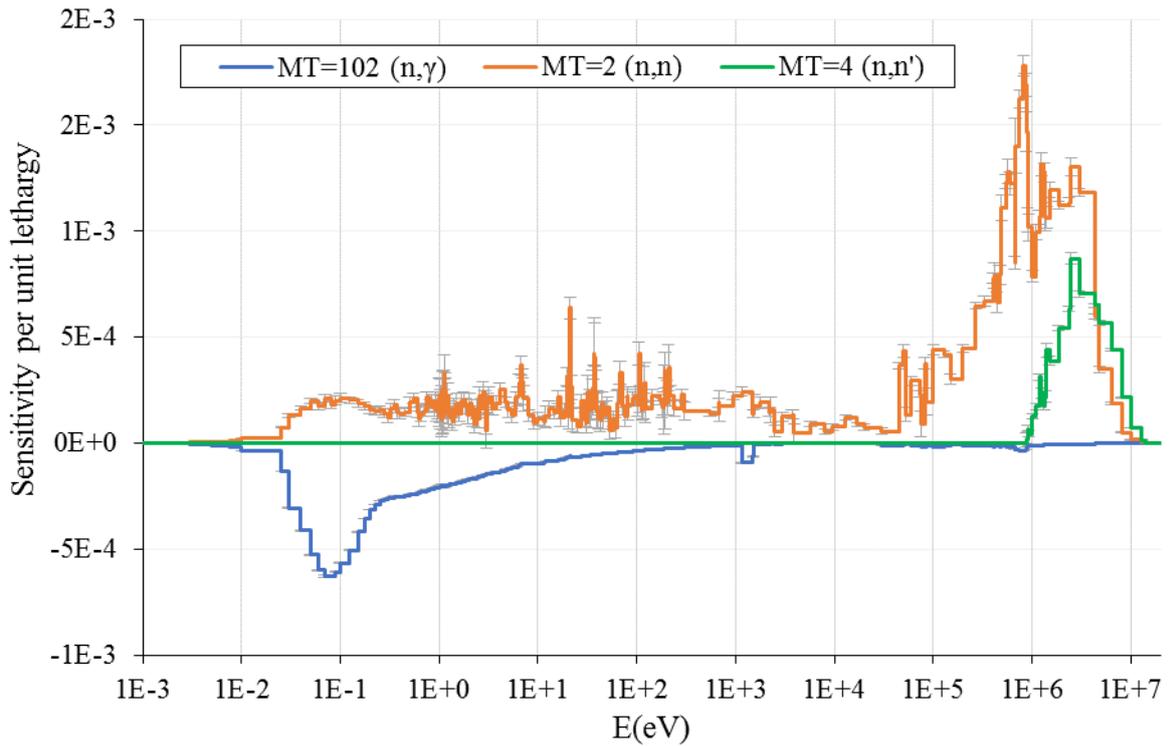

Figure 5: Sensitivity for $^{56}$Fe in baffle (in Case 3)

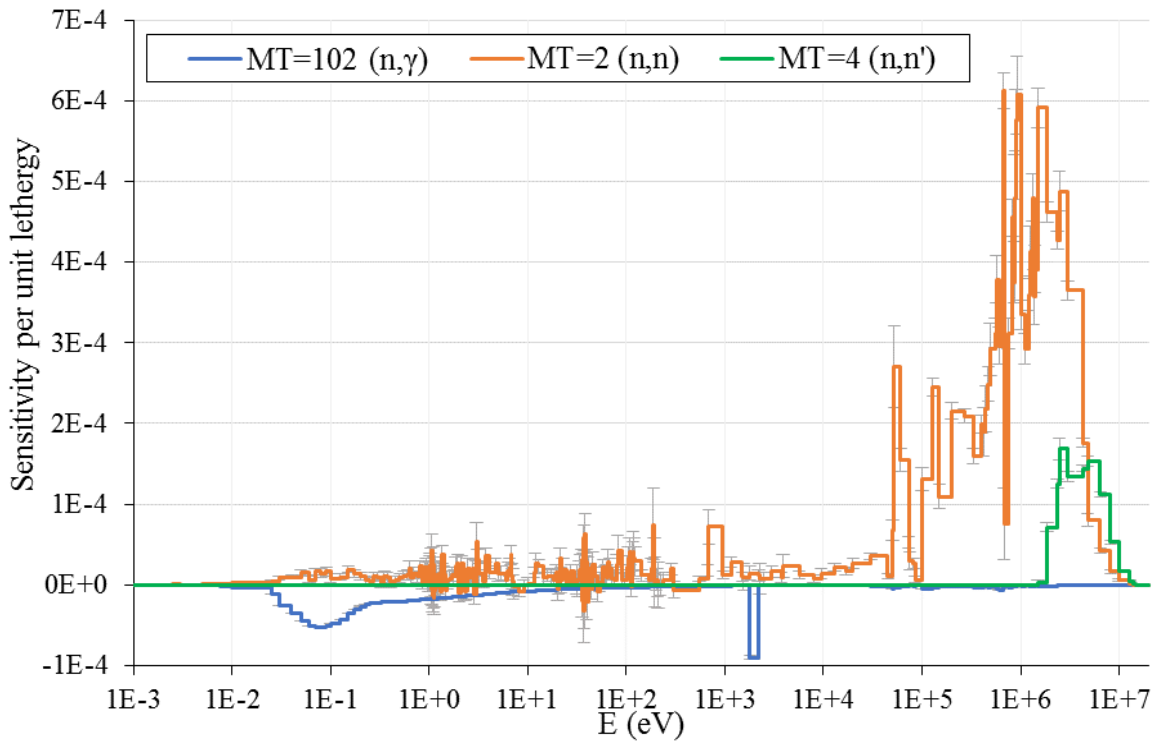

Figure 6: Sensitivity for $^{52}$Cr in baffle (in Case 3)

### 3.3 Fission density profile

Based on the measured NPA at selected positions in the irradiated fuel pins, the fission density was determined (see Figure 7). The fission densities were normalized to the inner pins (positions

17-24), where the shape of the neutron flux profile is identical in all cases. As expected, in the part affected by various reflectors, the lowest flux is in the non-reflected case, while the highest one in the case of baffle where the two cooling channels are filled by steel (Case 4). The profile in the water reflected case (Case 1) is not symmetrical due to the presence of instrumentation channel close to nether pins in the cut across the core (Figure 2).

The measured fission density profile was compared with calculations (see Figure 8). One can say that the agreement is satisfactory, being within 4 % in most of the cases. The outliers are close to water air boundaries and reflect the uncertainty in water thickness between fuel pin and dry assembly. Generally, the uncertainties in C/E comparison in non-boundary pins are about 2-3%, and in boundary pins are about 5%.

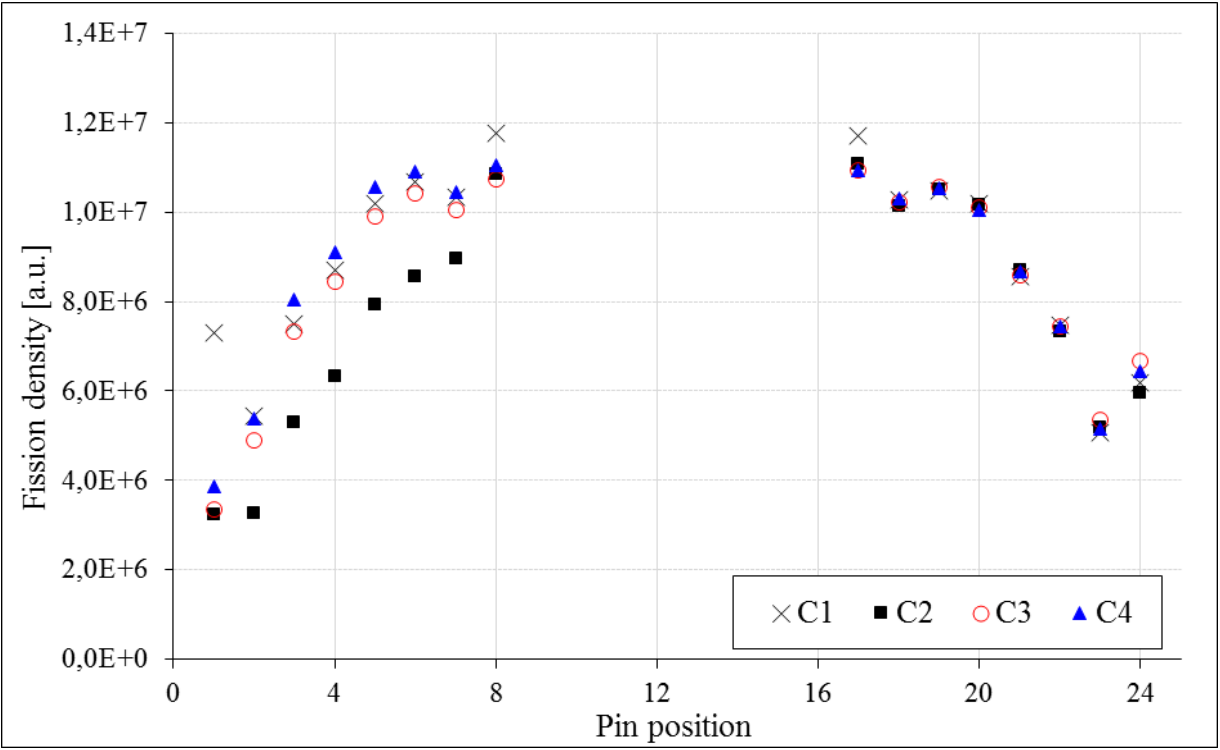

Figure 7: Measured fission density profile for various cores

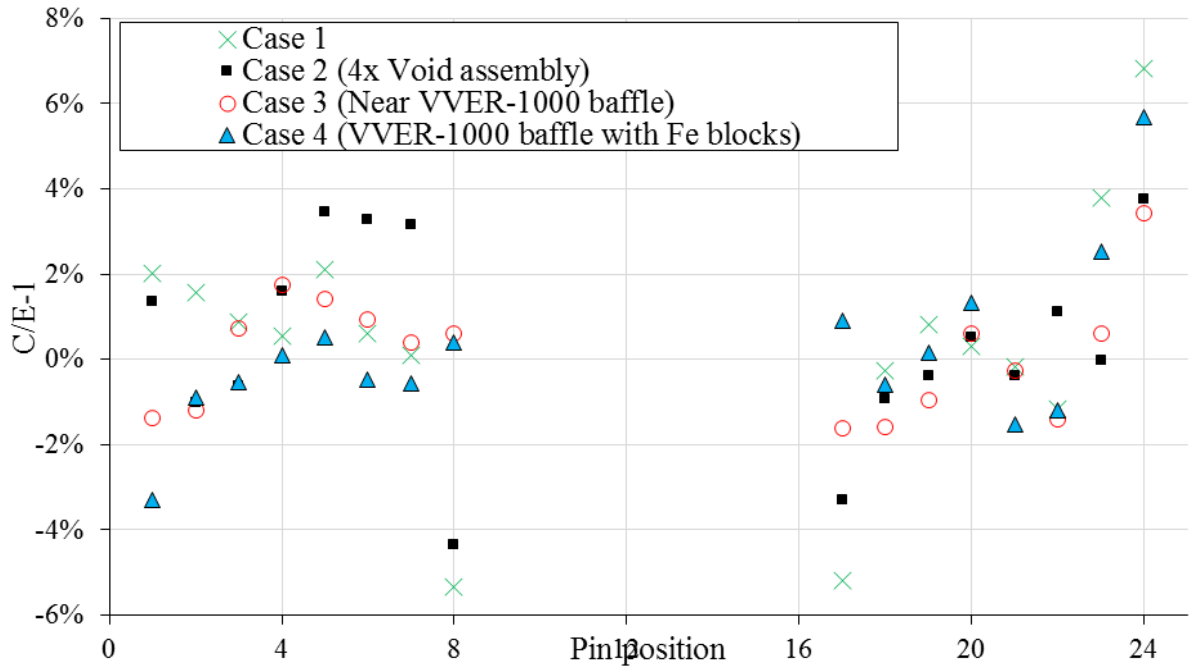

Figure 8: C/E-1 agreement of fission density across the studied cores

Extrapolation lengths for each reflector case were derived from the experimentally measured fission densities. The results (listed in Table 8) represent the correlation with experimentally determined critical water levels of each criticality measurement. The highest extrapolation length in Case 4 is the response on the best neutron economy and the lowest critical water level. The second best neutron economy shows the water reflector in Case 1. The VVER-1000 (Case 3) has a shorter extrapolation length than Case 1.

Table 8.: Experimentally determined extrapolated length

| Case | $L_{extr.}$ |
|---|---|
| Case 1 | 50.09 |
| Case 2 | 42.71 |
| Case 3 | 49.83 |
| Case 4 | 53.1 |

As the fission density profile was evaluated only in a relative way, thus the question of agreement in absolute comparison arises. Due to this fact, an independent fuel irradiation experiment was carried out simultaneously with the irradiation of pure Mn dosimeters to normalize the power in the positions where the reference neutron field was defined. The geometry in this case corresponded to Case 1A. A good agreement of the calculated and experimentally measured fission densities was obtained (see Figure 9). Experimental fission densities normalized per 1 core neutron are listed in Table 9.

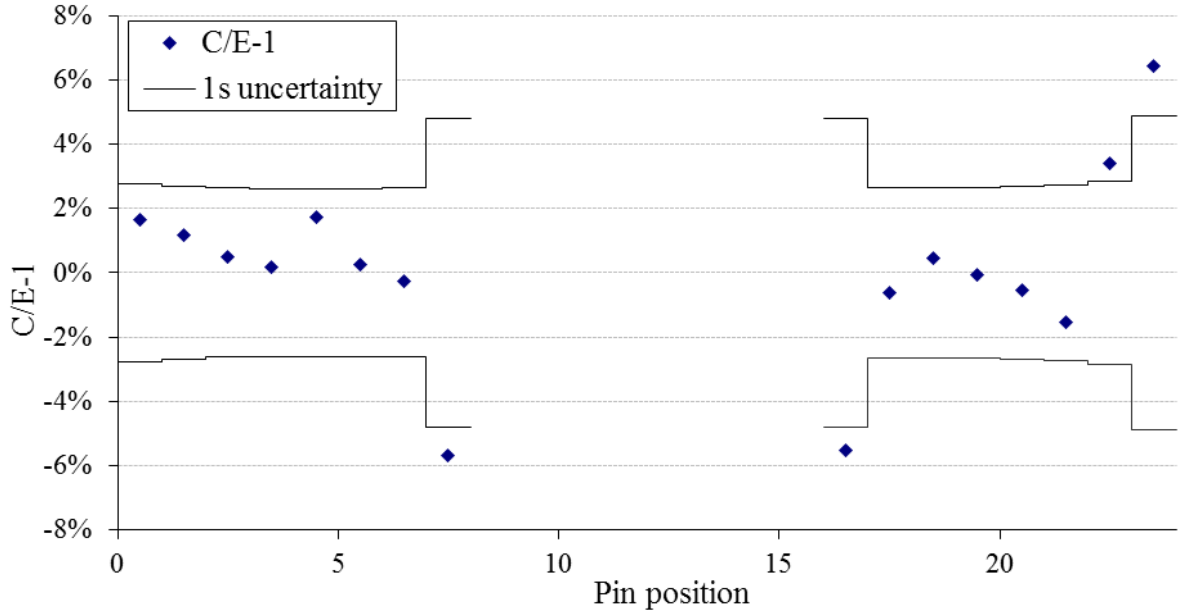

Figure 9: C/E-1 of fission density determined across Case 1A core

Table 9.: Fission density distribution across core per 1 neutron, (Mn foils used in normalization)

| Pin position | Row position | Y [cm] | Fission density [fiss/s/nps] | Rel. unc. |
|---|---|---|---|---|
| 1 | 1 | 34.64 | 1.81E-04 | 2.8% |
| 2 | 3 | 32.43 | 1.35E-04 | 2.7% |
| 3 | 7 | 28.02 | 1.85E-04 | 2.6% |
| 4 | 9 | 25.81 | 2.15E-04 | 2.6% |
| 5 | 13 | 21.39 | 2.52E-04 | 2.6% |
| 6 | 15 | 19.18 | 2.64E-04 | 2.6% |
| 7 | 19 | 14.77 | 2.55E-04 | 2.6% |
| 8 | 21 | 12.56 | 2.91E-04 | 4.8% |
| 17 | 43 | -12.56 | 2.89E-04 | 4.8% |
| 18 | 45 | -14.77 | 2.54E-04 | 2.7% |
| 19 | 49 | -19.18 | 2.59E-04 | 2.6% |
| 20 | 51 | -21.39 | 2.52E-04 | 2.7% |
| 21 | 55 | -25.81 | 2.12E-04 | 2.7% |
| 22 | 57 | -28.02 | 1.84E-04 | 2.7% |
| 23 | 61 | -32.43 | 1.25E-04 | 2.9% |
| 24 | 63 | -34.64 | 1.53E-04 | 4.9% |

### 3.4 Neutron flux profile

The effect of the significant reactivity worth when the water column is replaced by some steel has a relatively simple explanation. It is caused by the higher fast neutron flux in the core and baffle boundary (see Figure 12), which is also affecting the surrounding fuel. This higher fast neutron flux is the result of better fast neutron backscattering properties of steel compared to water. As the majority of fission is caused by thermal neutrons, and an increase in the flux of fast neutrons is observed, an increase in the fission rate will manifest itself when these neutrons are moderated, which occurs after the first pin row.

The local power increase in Case 1 (with a water reflector) in the first row of fuel pins, which is greater than in Case 3 or 4 (VVER-1000 reflector), is caused by a significantly better thermal neutron albedo of water than of steel.

The two peaks in the calculated thermal neutron flux profile across internals in Case 4 are the result of the moderating effect of water in the gaps between the steel insertion of diameter 12.5 cm and cooling channel in the VVER-1000 internals of diameter 13 cm. This effect has only a local effect due to the high absorption of thermal neutrons in the baffle.

Similarly, the case where the cooling channel is filled with steel leads to a high increase of the fast neutron flux at the baffle core boundary, which highly increases the fast flux in fuel. As the thermal neutron flux is comparable in both cases, the change of water level between Case 3 and Case 4 is more evident than between Case 1A and Case 4. The plot with the relative change in power profile is in Figure 10.

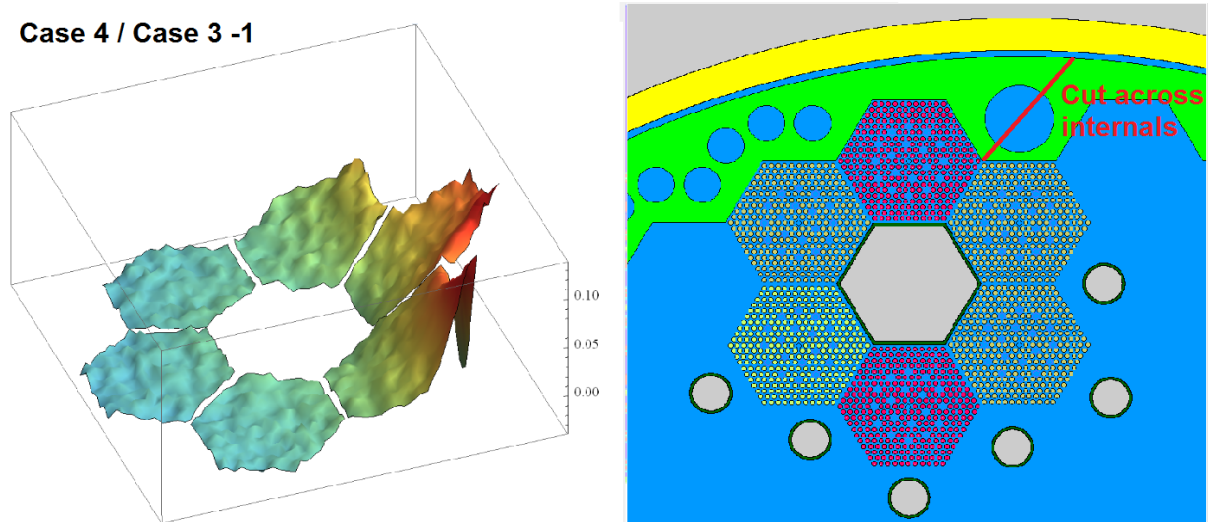

Figure 10: Change in power profile when steel block inserted into internals (left). Cut across internals where the flux profile was determined (right) see Figure 11 and Figure 12

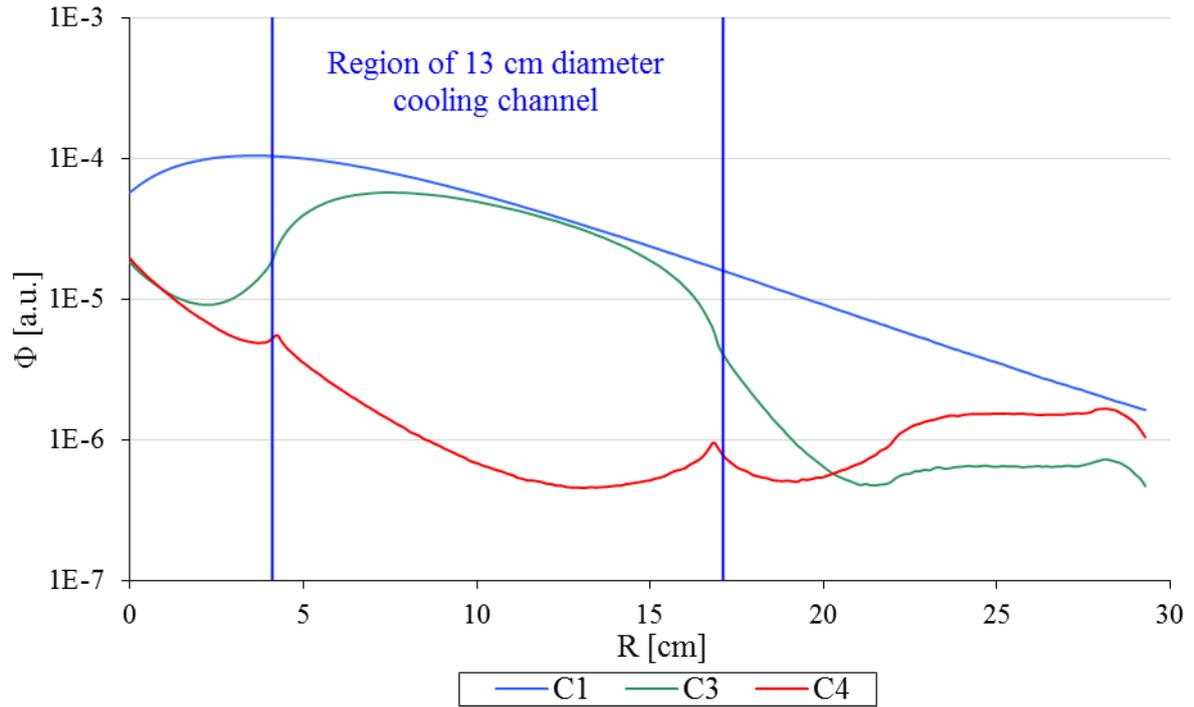

Figure 11: Calculated profile of thermal neutron flux across internals

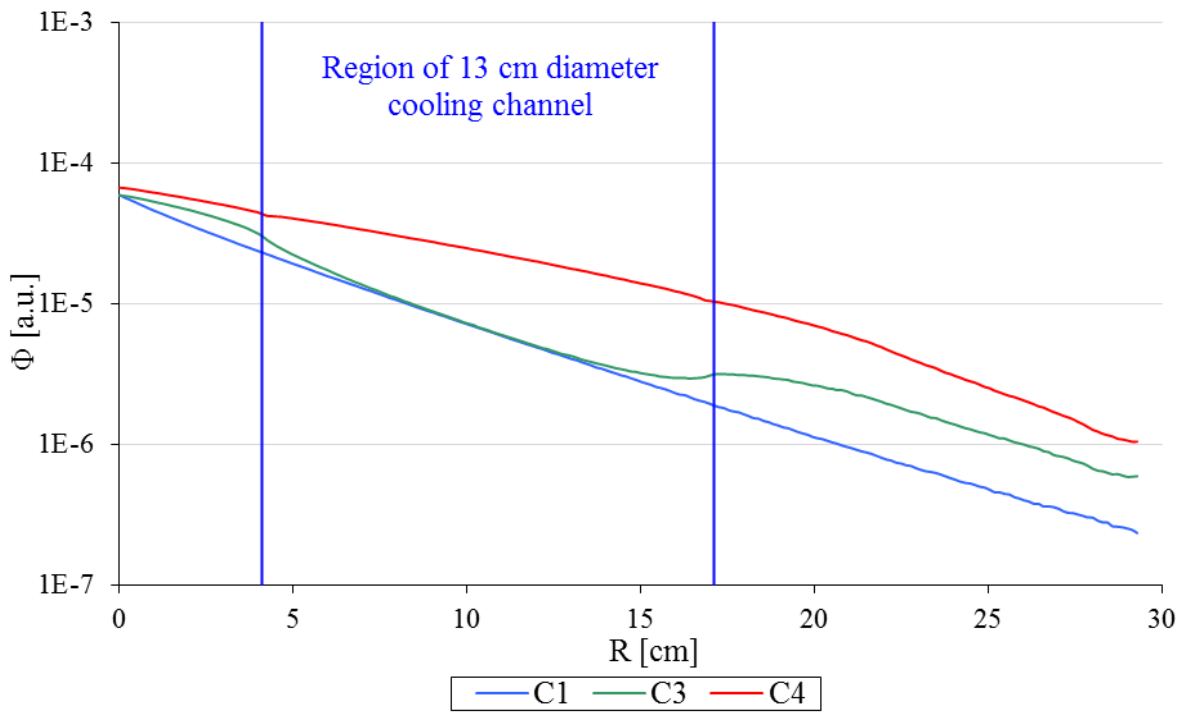

Figure 12: Calculated profile of fast neutron flux > 0.1 MeV across internals

## 4 Conclusions

It has been shown that the cooling channels in VVER-1000 reactor internals have a strong negative effect on the core neutron economy. This experimentally obtained result is confirmed by calculation. On the other hand, the presence of cooling channels affects the temperature field, so that the neutron economy is balanced by suppressing the effect of void swelling. According to the previous results, it was obtained a good agreement between the calculation and the experiment in the case of the water reflector. A similar situation was observed for the partially non reflected core. The rate of agreement significantly worsens when the core is next to the VVER-1000 reflector. A slightly better agreement was obtained with the case where the two most effective cooling channels were filled by steel inserts. Calculated criticality shows weak sensitivity to the undertaken changes in new INDEN iron and chromium evaluations.

## 5 Acknowledgements

Presented results were obtained with the use of the infrastructure Reactors LVR-15 and LR-0, which is financially supported by the Ministry of Education, Youth and Sports - project LM2015074, the SANDA project funded under H2020-EURATOM-1.1 contract 847552.